\documentclass[prd,twocolumn,showpacs,floatfix,amsmath,nofootinbib,amssymb,floatfix]{revtex4}
\usepackage{graphicx,color,dcolumn,booktabs,bm}
\usepackage{longtable,lscape}
\usepackage{txfonts}
\usepackage{overpic}
\usepackage{amssymb}
\usepackage{indentfirst}
\usepackage{feynmf}   
\usepackage{slashed}  
\usepackage{cases}
\usepackage{color}
\usepackage{multirow}
\usepackage{epstopdf}
\usepackage{longtable}
\usepackage{graphicx,color,dcolumn,booktabs,bm}
\usepackage[colorlinks,
            citecolor=blue,
            anchorcolor=red,
            menucolor=red,
            linkcolor=red,
            filecolor=red,
            runcolor=red,
            urlcolor=blue,
            frenchlinks=red]{hyperref}

\graphicspath{{Figures/}} %

\allowdisplaybreaks

\begin{document}

\title{From the isovector molecular explanation of the newly $T_{c\bar{s}}^{a0(++)}(2900)$ to possible charmed-strange molecular pentaquarks}

\author{Rui Chen$^{1}$}\email{chenrui@hunnu.edu.cn}
\author{Qi Huang$^{2}$\footnote{Corresponding author}}\email{huangqi@ucas.ac.cn}

\affiliation{
$^1$Key Laboratory of Low-Dimensional Quantum Structures and Quantum Control of Ministry of Education, Department of Physics and Synergetic Innovation Center for Quantum Effects and Applications, Hunan Normal University, Changsha 410081, China\\
$^2$Department of Physics, Nanjing Normal University, Nanjing 210023, China}
\date{\today}

\begin{abstract}
In this work, we adopt the one-boson-exchange model to study the $D^{(*)}K^{(*)}$ interactions. With the same parameters, we can simultaneously reproduce the masses of the $D_{s0}(2317)$, $D_{s1}(2460)$, and $T_{c\bar{s}}^{a0(++)}(2900)$ recently observed by the LHCb collaboration in the hadronic molecular picture, where the $D_{s0}(2317)$, $D_{s1}(2460)$, and $T_{c\bar{s}}^{a0(++)}(2900)$ are regarded as the $DK[I(J^P)=0(0^+)]$, $D^*K[0(1^+)]$, and $D^*K^*[1(0^+)]$ charmed-strange molecular states, respectively. In addition, we extend to study the $\Lambda_cK^{(*)}$ and $\Sigma_cK^{(*)}$ interactions and predict two possible charmed-strange molecular pentaquarks, the single $\Sigma_cK^*$ state with $I(J^P)=1/2(1/2^-)$ and $3/2(3/2^-)$. After considering the coupled channel effects, our results show that the coupled $\Lambda_cK^*/\Sigma_cK^*$ molecular state with $I(J^P)=1/2(1/2^-)$ and the coupled $\Sigma_cK/\Lambda_cK^*$ molecular state with $I(J^P)=1/2(1/2^-)$ can be good hadronic molecular candidates, the $\Lambda_cK^*({}^2S_{1/2})$ and $\Sigma_cK({}^2S_{1/2})$ are the dominant channels, respectively.

\end{abstract}


\maketitle

\section{introduction}

Apart from the conventional mesons and baryons given at the birth of the quark model \cite{Gell-Mann:1964ewy}, the existence of the deuteron naturally indicates that hadronic molecules must also be QCD allowed hadronic states. Especially, with the discoveries of a serial of $XYZ$ states, the hadronic molecules are proved to play important roles in explaining many of them (one can see, for example, Refs. \cite{Chen:2016qju,Chen:2016spr,Guo:2017jvc,Chen:2022asf} for a detailed review).

There may have mainly two reasons why hadronic molecular interpretations are often introduced in interpreting these observed $XYZ$ states. One is that some of their properties cannot be explained by the conventional mesons or baryons. The other, which may be more important, is that their masses are just serval to serval tens MeVs below some specific thresholds.

In showing the above two reasons, a very good example is the newly discovered $T_{cc}^+(3875)$ \cite{LHCb:2021vvq,LHCb:2021auc}. Just in the last year, the LHCb collaboration reported this very narrow structure in the $D^0 D^0 \pi^+$ spectrum \cite{LHCb:2021vvq,LHCb:2021auc}, whose peak position from the $D^{\ast +}D^0$ threshold and width are $-273 \pm 61 \pm 5 ^{+11}_{-14}$ keV and $410 \pm 165 \pm 43 ^{+18}_{-38}$ keV respectively. \cite{LHCb:2021vvq,LHCb:2021auc}. From its mass, charged property and decay channel ($D^0 D^0 \pi^+$), it is easy to infer that if this state really exists, its minimal valance quark components must be $cc\bar{u}\bar{d}$, which means it cannot be collected into conventional mesons ($q\bar{q}$). Secondly, one can easily see that its mass is just about 300 keV below the $D^{\ast+}D^0$ threshold, which makes it be a very good candidate as a very loosely bound $D^{\ast+}D^0$ molecular state \cite{Janc:2004qn,Li:2012ss,Li:2021zbw,Ohkoda:2012hv,Liu:2019stu,Ding:2020dio,Chen:2021vhg,Feijoo:2021ppq,Wang:2021yld,Wang:2021ajy,Ren:2021dsi,Deng:2021gnb,Albaladejo:2021vln,Dai:2021vgf,Dai:2022ulk,Du:2021zzh,Meng:2021jnw,Ling:2021bir,Yan:2021wdl,Fleming:2021wmk,Huang:2021urd,Abreu:2022lfy,He:2022rta,Xin:2021wcr,Agaev:2022ast,Ke:2021rxd}.

The stories of the observations on the exotic states and the hadronic molecules still go on. Very recently, the LHCb collaboration continued to report their new observations of another two states $T_{c\bar{s}}^{a0}(2900)$ and $T_{c\bar{s}}^{a++}(2900)$, where the superscript $a$ means that their quantum numbers are both $I(J^P)=1(0^+)$ \cite{LHCb:Tcc,LHCb:Qian}. For the $T_{c\bar{s}}^{a0}(2900)$, the channel it get discovered is $D_s^+ \pi^-$, and its mass and width are $2892 \pm 14 \pm 15$ MeV and $119 \pm 26 \pm 12$ MeV, respectively \cite{LHCb:Tcc,LHCb:Qian}, while for the $T_{c\bar{s}}^{a++}(2900)$, these are $D_s^+ \pi^+$, $2921 \pm 17 \pm 19$ MeV and $137 \pm 32 \pm 14$ MeV, respectively \cite{LHCb:Tcc,LHCb:Qian}.

According to their decay channels, mass positions and quantum numbers, it is easy to guess that the $T_{c\bar{s}}^{a0}(2900)$ and $T_{c\bar{s}}^{a++}(2900)$ may belong to the same isovector triplet. In addition, due to the charged property of $T_{c\bar{s}}^{a0(++)}(2900)$, their minimal valance quark components are naturally inferred to be $c\bar{s}q\bar{q}$ ($q=u,~d$). However, for this quark components, it is not the first time that it shows up since there already exists two very famous candidates, $D_{s0}^\ast(2317)$ and $D_{s1}(2460)$.

The $D_{s0}^\ast(2317)$ was firstly observed by the BaBar collaboration in the $D_s^+ \pi^0$ invariant mass spectrum in the $B$ decay process \cite{BaBar:2003oey}, whose mass is about 2.32 GeV. Later, in confirming this state, the CLEO collaboration found another state $D_{s1}(2460)$ \cite{CLEO:2003ggt}. After that, these two states are also confirmed by the Belle and BaBar experiments \cite{Belle:2003kup,BaBar:2006eep,BaBar:2003cdx}. Since the mass positions of these two states are far from the theoretical predictions of the masses of the charmed-strange mesons in the $(0^+,1^+)$ doublet, apart from the conventional meson interpretations, various exotic assignments were proposed, where hadronic molecules are included (one can refer to, for instance, Refs. \cite{Chen:2016spr,Guo:2017jvc} for a detailed review about the different assignments of these two states). Under the molecular scenario, $D_{s0}^\ast(2317)$ and $D_{s1}(2460)$ are interpreted as the isoscalar $S$-wave $DK$ and $D^\ast K$ molecule states respectively, and one important reason for this explanation is that they are just below the $DK$ and $D^\ast K$ thresholds, and the mass difference between these two states is equal to the energy difference between the $DK$ and $D^\ast K$ thresholds.

Now we look back on the newly observed $T_{c\bar{s}}^{a0}(2900)$ and $T_{c\bar{s}}^{a++}(2900)$. From their mass positions, it is easy to see that they are just near the $D^\ast K^\ast$ threshold. Since the channel they get observed is also $D_s \pi$, it is so hard to rule out their relationships with $D_{s0}^\ast(2317)$ and $D_{s1}(2460)$. In addition, from their mass positions, we can see that their mass differences with $D_{s0}^\ast(2317)/D_{s1}(2460)$ are approximately equal to the difference between $D^\ast K^\ast$ and $DK/D^\ast K$ thresholds. Thus, in our view, studying the possibility that these two new states being candidates of $D^\ast K^\ast$ molecules is not meaningless. Since the corresponding $D^{(*)} K^{(*)}$ interactions can be quantitatively described by the one-boson-exchange (OBE) model, with the derived effective potentials, we can find possible bound state solutions by solving the coupled channel Schr\"{o}dinger equations, and check the possibility of the newly $T_{c\bar{s}}^{a0(++)}(2900)$ as the isovector $D^\ast K^\ast$ molecule with $J^P=0^+$.

As byproducts, we further study the interactions between a charmed baryon $(\Lambda_c,~\Sigma_c)$ and a strange meson $(K^{(*)})$ to search for the possible charmed-strange molecular pentaquarks. In the future, experimental examinations of our predictions will not only enrich the family of the exotic states, but also help us to understand the nature of the newly $T_{c\bar{s}}^{a0(++)}(2900)$ together with the $D_{s0}^\ast(2317)$ and $D_{s1}(2460)$ in a sense.

This paper is organized as follows. After this introduction, we deduce the $D^{(*)}K^{(*)}$ interactions by using the OBE model and illustrate why the $T_{c\bar{s}}^{a0(++)}(2900)$ can be the isovector $D^\ast K^\ast$ molecule with $J^P=0^+$ in Sec.~\ref{sec2}. In Sec.~\ref{sec3}, we predict several possible charmed-strange molecular pentaquarks from the $\Lambda_cK^{(*)}$ and $\Sigma_cK^{(*)}$ interactions. The paper ends with a summary in Sec. \ref{sec4}.

\section{The $T_{c\bar s}^{a0(++)}$ as the isovector $D^*K^*$ molecular state with $J^P=0^+$}\label{sec2}

Before deducing the OBE effective potentials, we first introduce the relevant effective Lagrangians, for the charmed meson sector, the effective Lagrangians are constructed according to the heavy quark symmetry and chiral symmetry \cite{Yan:1992gz,Wise:1992hn,Burdman:1992gh,Casalbuoni:1996pg,Falk:1992cx},
\begin{eqnarray}
\mathcal{L}_{{\mathcal{P}}^{(*)}{\mathcal{P}}^{(*)}\sigma} &=& -2g_s{\mathcal{P}}_b^{\dag}{\mathcal{P}}_b\sigma-2g_s{\mathcal{P}}_b^{*}\cdot{\mathcal{P}}_b^{*\dag}\sigma,\label{lag1}\\
\mathcal{L}_{{\mathcal{P}}^{(*)}{\mathcal{P}}^{(*)}{P}} &=&-i\frac{2g}{f_{\pi}}v^{\alpha}\varepsilon_{\alpha\mu\nu\lambda}{\mathcal{P}}_b^{*\mu}{\mathcal{P}}_{a}^{*\lambda\dag}
\partial^{\nu}{P}_{ba},\\
\mathcal{L}_{{\mathcal{P}}^{(*)}{\mathcal{P}}^{(*)}{V}} &=& -\sqrt{2}\beta g_V{\mathcal{P}}_b{\mathcal{P}}_a^{\dag} v\cdot{V}_{ba}\nonumber\\
    &&+\sqrt{2}\beta g_V{\mathcal{P}}_b^*\cdot{\mathcal{P}}_a^{*\dag}v\cdot{V}_{ba}\nonumber\\
   &&-i2\sqrt{2}\lambda g_V{\mathcal{P}}_b^{*\mu}{\mathcal{P}}_a^{*\nu\dag}
   \left(\partial_{\mu}{V}_{\nu}-\partial_{\nu}{V}_{\mu}\right)_{ba}.
\end{eqnarray}
Here, $v=(1,\textbf{0})$, $\mathcal{P}^{(*)}$ stands for the pseudoscalar (or vector) mesons fields $\mathcal{P}^{(*)T}=(D^{(*)+},~D^{(*)0})$.
$\mathcal{V} = \frac{1}{2}(\xi^{\dag}\partial_{\mu}\xi+\xi\partial_{\mu}\xi^{\dag})$ and $A_{\mu} = \frac{1}{2}(\xi^{\dag}\partial_{\mu}\xi-\xi\partial_{\mu}\xi^{\dag})$ are the vector current and axial current, respectively. $\xi=\text{exp}(i{P}/f_{\pi})$, $f_{\pi}=132$ MeV.
$F_{\mu\nu}(\rho)=\partial_{\mu}\rho_{\nu}-\partial_{\nu}\rho_{\mu}+[\rho_{\mu},\rho_{\nu}]$ with $\rho_{\mu}=ig_{V}{V}_{\mu}/\sqrt{2}$. $g=0.59\pm 0.07\pm 0.01$ is extracted from the decay width of $D^*\to D\pi$ \cite{Casalbuoni:1996pg}. $\beta=$0.9 is fixed by the vector meson dominance \cite{Isola:2003fh}. $\lambda=$ 0.56 GeV$^{-1}$ is determined through a comparison of the form factor between the theoretical calculation from the light cone sum rule and lattice QCD \cite{Isola:2003fh}. $g_V=m_{\rho}/f_{\pi}=5.8$. Matrices ${P}$ and ${V}$ are written as
\begin{eqnarray}
{P} &=& \left(\begin{array}{cc}
\frac{\pi^0}{\sqrt{2}}+\frac{\eta}{\sqrt{6}} &\pi^+  \nonumber\\
\pi^- &-\frac{\pi^0}{\sqrt{2}}+\frac{\eta}{\sqrt{6}}
\end{array}\right),\nonumber\\
{V} &=& \left(\begin{array}{cc}
\frac{\rho^0}{\sqrt{2}}+\frac{\omega}{\sqrt{2}} &\rho^+  \nonumber\\
\rho^- &-\frac{\rho^0}{\sqrt{2}}+\frac{\omega}{\sqrt{2}}
\end{array}\right).\label{lag3}
\end{eqnarray}

The effective Lagrangians describing the interactions between the strange mesons and light mesons are constructed in the $SU(3)$ symmetry \cite{Lin:1999ad,Nagahiro:2008mn}, which read as
\begin{eqnarray}
\mathcal{L}_{K^{(*)}K^{(*)}\sigma} &=& g_{\sigma }m_K\bar{K} K\sigma-g_{\sigma }m_{K^*}\bar{K}^{*}\cdot K^{*}\sigma,\\
\mathcal {L}_{\rho KK} &=& ig_{\rho KK}\left(\bar K\partial^{\mu}K
       - \partial^{\mu}\bar{K}K\right)\bm{\tau}\cdot\bm{\rho_{\mu}},\\
\mathcal {L}_{\omega KK} &=& ig_{\omega KK}\left(\bar K\partial^{\mu}K
       - \partial^{\mu}\bar{K}K\right)\omega_{\mu},\\
\mathcal {L}_{\rho K^*K^*} &=&
       ig_{\rho K^*K^*}\left[(K^*_{\nu}\partial^{\mu}\bar{K}^{{*\nu}}
       -\partial^{\mu}K^{*\nu}\bar{K}^{*}_{\nu})\bm{\tau}\cdot\bm{\rho_{\mu}}\right.\nonumber\\
       &&+(\partial^{\mu}K^{*\nu}\bar{K}^{*}_{\mu}
       -K^*_{\mu}\partial^{\mu}\bar{K}^{*\nu})\bm{\tau}\cdot\bm{\rho_{\nu}}\nonumber\\
       &&+\left.(K^*_{\mu}\bar{K}^{*}_{\nu}-K^*_{\nu}\bar{K}^{*}_{\mu})
       \bm{\tau}\cdot\partial^{\mu}\bm{\rho^{\nu}}\right],\\
\mathcal {L}_{\omega K^*K^*} &=& ig_{\omega K^*K^*}
       \left[(K^*_{\nu}\partial^{\mu}\bar{K}^{{*\nu}}
       -\partial^{\mu}K^{*\nu}\bar{K}^{*}_{\nu})\omega_{\mu}\right.\nonumber\\
       &&+(\partial^{\mu}K^{*\nu}\bar{K}^{*}_{\mu}
       -K^*_{\mu}\partial^{\mu}\bar{K}^{*\nu})\omega_{\nu}\nonumber\\
       &&+\left.(K^*_{\mu}\bar{K}^{*}_{\nu}-K^*_{\nu}\bar{K}^{*}_{\mu})\partial^{\mu}\omega^{\nu}\right],\\
\mathcal {L}_{\pi K^*K^*} &=& -g_{\pi K^*K^*}\varepsilon^{\mu\nu\rho\sigma}
        \partial_{\mu}K^*_{\nu}\partial_{\rho}\bar{K}^{*}_{\sigma}\bm{\tau}\cdot\bm{\pi},\\
\mathcal {L}_{\eta K^*K^*} &=& g_{\eta K^*K^*}\varepsilon^{\mu\nu\rho\sigma}
        \partial_{\mu}K^*_{\nu}\partial_{\rho}\bar{K}^{*}_{\sigma}\eta.\label{lag8}
\end{eqnarray}
In the above Lagrangians, we estimate the $\sigma$ meson couplings by the quark model as $g_s=g_{\sigma}=g_{\sigma NN}/3$ with ${g_{\sigma NN}^2}/({4\pi})=5.69$ \cite{Wang:2019aoc}. In Ref. \cite{Chen:2011cj}, $g_{\rho {K}^{(*)}{K}^{(*)}} =G/4 =3.425$, $g_{\omega {K}^{(*)}{K}^{(*)}} = 4.396$. $g_{\pi {K}^*{K}^*}$ and $g_{\eta {K}^*{K}^*}$ are taken as $g_{\pi {K}^*{K}^*}  =\frac{G^2N_c}{64\pi^2f_{\pi}}$, and
$g_{\eta {K}^*{K}^*}   =\frac{G^2N_c}{64\sqrt{3}\pi^2f_{\pi}}$ \cite{Kaymakcalan:1983qq}, $N_c$ is the number of color.

With these prepared effective Lagrangians, we can easily write down the scattering amplitudes for the $D^{(*)}K^{(*)}\to D^{(*)}K^{(*)}$ processes in the $t-$channel. The corresponding effective potentials can be related to the scattering amplitude by the Breit approximation,
\begin{eqnarray}\label{breit}
\mathcal{V}_{E}^{D^{(*)}K^{(*)}\to D^{(*)}K^{(*)}}(\bm{q}) &=&
          -\frac{\mathcal{M}(D^{(*)}K^{(*)}\to D^{(*)}K^{(*)})}
          {\sqrt{\prod_i2M_i\prod_f2M_f}}.
\end{eqnarray}
Here, $M_i$ and $M_f$ are the masses of the initial states and final states, respectively. $\mathcal{M}(D^{(*)}K^{(*)}\to D^{(*)}K^{(*)})$ denotes the scattering amplitude for the $D^{(*)}K^{(*)}\to D^{(*)}K^{(*)}$ process by exchanging the light mesons ($\sigma$, $\pi$, $\eta$, $\rho$, and $\omega$). Next, we perform the Fourier transformation to obtain the effective potentials in the coordinate space $\mathcal{V}(\bm{r})$,
\begin{eqnarray}
\mathcal{V}_{E}(\bm{r}) =
          \int\frac{d^3\bm{q}}{(2\pi)^3}e^{i\bm{q}\cdot\bm{r}}
          \mathcal{V}_{E}(\bm{q})\mathcal{F}^2(q^2,m_E^2).\nonumber
\end{eqnarray}
In order to compensate the off-shell effect of the exchanged meson, we introduce a monopole form factor $\mathcal{F}(q^2,m_E^2)= (\Lambda^2-m_E^2)/(\Lambda^2-q^2)$ at every interactive vertex, where $m_E$ and $q$ are the mass and four-momentum of the exchanged meson, respectively.

The flavor wave functions $|I,I_3\rangle$ for the isovector $D^{(*)}K^{(*)}$ systems can be expressed as
\begin{eqnarray*}
|1,1\rangle &=& |K^{(*)+}D^{(*)+}\rangle,\\
|1,0\rangle &=& \frac{1}{\sqrt{2}}(|K^{(*)+}D^{(*)0}\rangle-|K^{(*)0}D^{(*)+}\rangle),\\
|1,-1\rangle &=& |K^{(*)0}D^{(*)0}\rangle,
\end{eqnarray*}
where $I$ and $I_3$ stand for the isospin and the third component, respectively. When we consider the $S-D$ wave mixing, the spin-orbit wave functions $|{}^{2S+1}L_J\rangle$ for the $D^{*}K^{*}$ system with $J^P=0^+$ include $|{}^{1}S_0\rangle$ and $|{}^{5}D_0\rangle$. The total OBE effective potentials for the isovector $D^{*}K^{*}$ systems can be written as
\begin{widetext}
\begin{eqnarray}
\mathcal{V}_{{D}^* K^*}(r)&=& -g_sg_{\sigma}\left(\begin{array}{cc}1 &0\\ 0 &1\end{array}\right)Y(\Lambda,m_{\sigma},r)
     +\frac{1}{6\sqrt{2}}\frac{gg_{\pi {K}^*{K}^*}}{f_{\pi}}
      \left[\left(\begin{array}{cc}2 &0\\ 0 &-1\end{array}\right)\nabla^2
      +\left(\begin{array}{cc}0 &\sqrt{2}\\ \sqrt{2} &2\end{array}\right)
      r\frac{\partial}{\partial r}\frac{1}{r}\frac{\partial}{\partial r}\right]
      \left(Y(\Lambda,m_{\pi},r)-\frac{1}{3}
      Y(\Lambda,m_{\eta},r)\right)\nonumber\\
      &&+\left[\frac{1}{2}\beta g_V\left(\begin{array}{cc}1 &0\\ 0 &1\end{array}\right)
      -\frac{1}{6}\frac{\lambda g_V}{m_{K^*}}
      \left(2\left(\begin{array}{cc}2 &0\\ 0 &-1\end{array}\right)\nabla^2
      -\left(\begin{array}{cc}0 &\sqrt{2}\\ \sqrt{2} &2\end{array}\right)
      r\frac{\partial}{\partial r}\frac{1}{r}\frac{\partial}{\partial r}\right)\right]
      \left(g_{\rho {K}^*{K}^*}Y(\Lambda,m_{\rho},r)-g_{\omega {K}^*{K}^*}Y(\Lambda,m_{\omega},r)\right),\label{dsks}
\end{eqnarray}
\end{widetext}
where $Y(\Lambda,m,{r}) = \frac{1}{4\pi r}(e^{-mr}-e^{-\Lambda
r})-\frac{\Lambda^2-m^2}{8\pi \Lambda}e^{-\Lambda r}$.

Before checking the possibility of the newly observed $T_{cs}(2900)$ as the isovetor $D^*K^*$ molecular state with $J^P=0^+$, we would like to estimate the cutoff value by reproducing the masses of the $D_{s0}(2317)$ and $D_{s1}(2460)$ as the $S-$wave isoscalar $DK$ and $D^*K$ molecular state, respectively. The flavor wave functions for the isoscalar $D^{(*)}K$ can be expressed as $|0,0\rangle=(|K^+D^{(*)0}\rangle+|K^0D^{(*)+}\rangle)/{\sqrt{2}}$. The OBE effective potentials for the isoscalar $DK$ system and the $D^*K$ system are
\begin{eqnarray}
\mathcal{V}_{DK }(r)&=& -g_sg_{\sigma}Y(\Lambda,m_{\sigma},r)
    -\frac{3}{2}\beta g_V g_{\rho KK}Y(\Lambda,m_{\rho},r)\nonumber\\
      &&-\frac{1}{2}\beta g_V g_{\omega KK} Y(\Lambda,m_{\omega},r),\label{ds0}\\
\mathcal{V}_{D^* K}(r) &=& -g_sg_{\sigma}(\bm{\epsilon}_1\cdot\bm{\epsilon}_3^{\dag})Y(\Lambda,m_{\sigma},r)\nonumber\\
     &&-\frac{3}{2}\beta g_V g_{\rho KK}(\bm{\epsilon}_1\cdot\bm{\epsilon}_3^{\dag})Y(\Lambda,m_{\rho},r)\nonumber\\
      &&-\frac{1}{2}\beta g_V g_{\omega KK}(\bm{\epsilon}_1\cdot\bm{\epsilon}_3^{\dag})Y(\Lambda,m_{\omega},r),\label{ds1}
\end{eqnarray}
respectively. For the $D^*K$ systems, if we consider the $S-D$ mixing effects, its spin-orbit wave functions include $|{}^3S_1\rangle$ and $|{}^3D_1\rangle$. In our numerical calculations, the operator $\bm{\epsilon}_1\cdot\bm{\epsilon}_3^{\dag}$ in Eq. (\ref{ds1}) will be replaced by the matrix elements ${diag}(1,1)$. Obviously, the $S-$wave OBE effective potentials for the isoscalar $DK$ and $D^*K$ systems are exactly the same as presented in Eqs. (\ref{ds0})-(\ref{ds1}).

\begin{figure}[!htbp]
\centering
\includegraphics[width=3.3in]{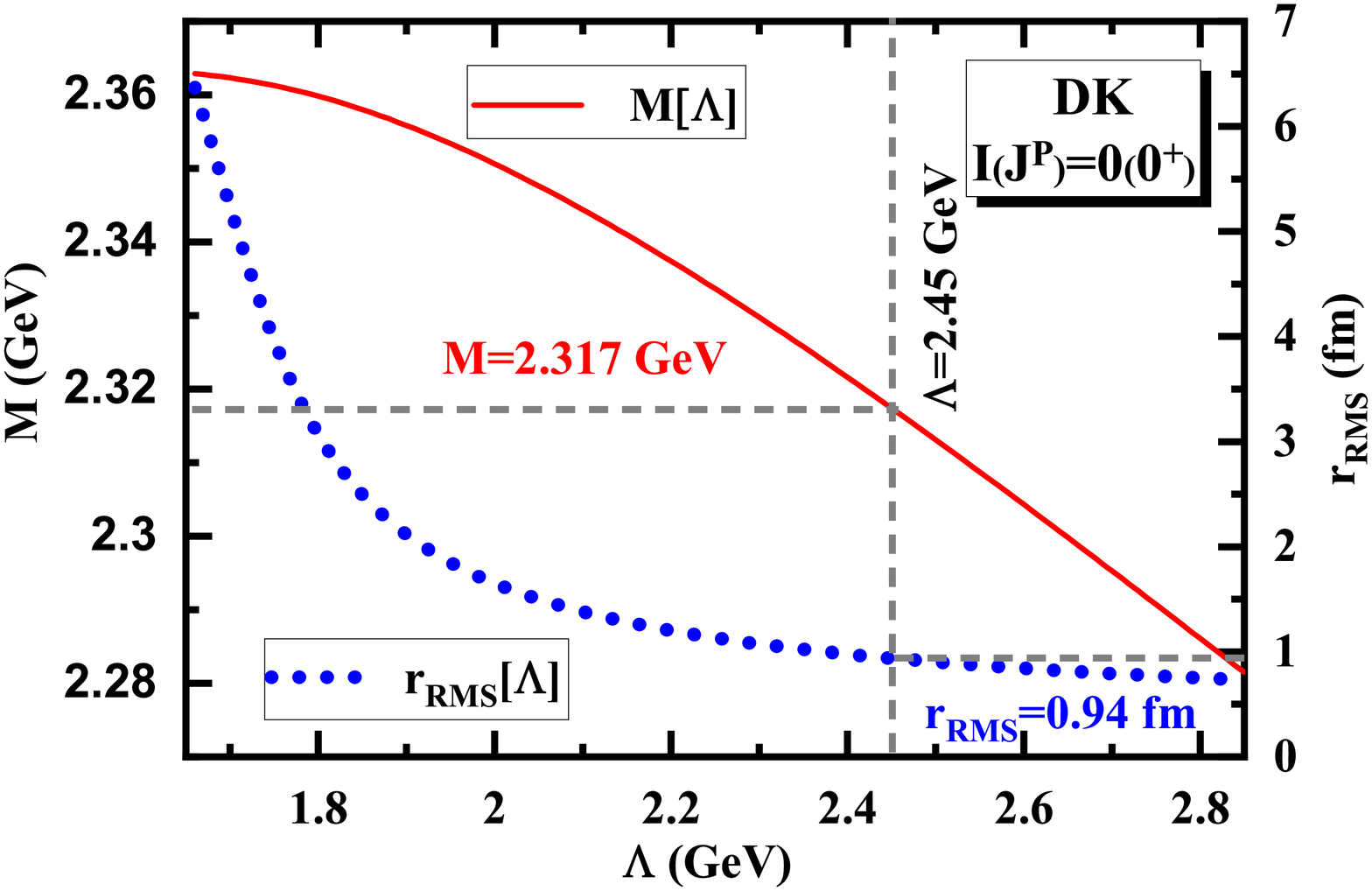}
\caption{The cutoff $\Lambda$ dependence of the bound solutions (mass $M$ and root-mean-square radius $r_{RMS}$) for the $DK$ system with $I(J^P)=0(0^+)$.}
\label{ds2317}
\end{figure}

In Fig. \ref{ds2317}, we present the cutoff $\Lambda$ dependence of the bound state solutions (mass and root-mean-square radius) for the $DK$ system with $I(J^P)=0(0^+)$. When the cutoff is taken as $\Lambda=2.45$ GeV, we obtain a bound state, its root-mean-square radius $r_{RMS}$ is 0.94 fm and its mass overlaps to the mass of the $D_{s0}(2317)$. If we adopt the same cutoff $\Lambda=2.45$ GeV in the $D^*K$ system with $I(J^P)=0(1^+)$, we can find a bound state with mass $M=2.456$ GeV and RMS radius $r_{RMS}=0.92$ fm, which can correspond to the $D_{s1}(2460)$. Thus, when the cutoff value is taken around 2.45 GeV, the $D_{s0}(2317)$ and $D_{s1}(2460)$ can be regarded as the isoscalar $DK$ molecular state with $J^P=0^+$ and the isoscalar $D^*K$ molecular state with $J^P=1^+$, respectively.

\begin{figure}[!htbp]
\centering
\includegraphics[width=3.3in]{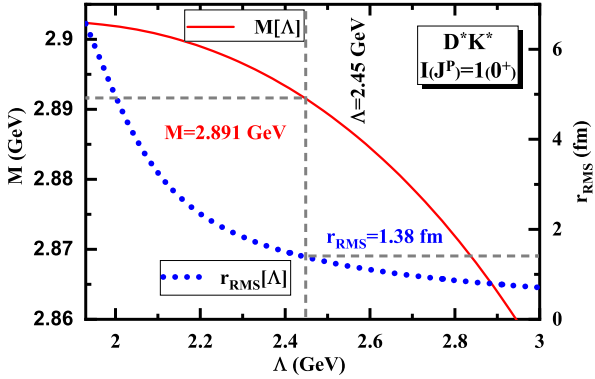}
\caption{The cutoff $\Lambda$ dependence of the bound solutions (mass $M$ and root-mean-square radius $r_{RMS}$) for the $D^*K^*$ system with $I(J^P)=1(0^+)$.}
\label{tcs2900}
\end{figure}

Now let's go to the $D^*K^*$ system with $I(J^P)=1(0^+)$. In Fig. \ref{tcs2900}, we present the cutoff $\Lambda$ dependence of the obtained mass and the RMS radius of the $D^\ast K^\ast$ system. Here, one can see when the cutoff is larger than 1.93 GeV, the binding energy is around several to several tens MeV, and the RMS radius is larger than or around 1 fm, which satisfy the typical properties of the loosely bound hadronic molecular state. If we still adopt the above cutoff $\Lambda=2.45$ GeV, the mass of the $D^*K^*$ molecule with $1(0^+)$ is 2.891 GeV, which overlaps to the mass of the newly observed $T_{c\bar{s}}^{a0(++)}(2900)$. Meanwhile, its RMS radius is 1.38 fm. In addition, we further explore the role of the $S-D$ wave mixing effects in binding this bound state, which turns out that the probabilities for the $S-$wave and $D-$wave components are 91.32\% and 8.68\%, respectively. If we only consider the $S-$wave interaction, the binding energies will become larger with the same cutoff input as summarized in Table \ref{num1}. For example, when the cutoff is $\Lambda=2.45$ GeV, the binding energy in the $S$ wave considered case is $-4.00$ MeV, it is larger than that $E=-10.96$ MeV in the $S-D$ wave mixing effects considered case. Therefore, the $S-D$ wave mixing effects are helpful in forming the $D^*K^*$ molecule with $1(0^+)$.

\renewcommand\tabcolsep{0.3cm}
\renewcommand{\arraystretch}{1.6}
\begin{table}[!hbtp]
\caption{The obtained bound-state solutions (binding energy $E$ and root-mean-square radius $r_{RMS}$) for isovector ${D}^{*} K^*$ system with $J^P=0^+$ with and without considering the $S-D$ wave mixing effects. Here, $E$, $r_{RMS}$, and $\Lambda$ are in units of MeV, fm, and GeV, respectively.}\label{num1}
\begin{tabular}{ccc|ccc}
\toprule[1pt]\toprule[1pt]
\multicolumn{3}{c|}{Only $S$ wave considered}    &\multicolumn{3}{c}{$S-D$ wave mixing considered}\\\midrule[1pt]
$\Lambda$   &$E$ &$r_{RMS}$
 &$\Lambda$    &$E$  &$r_{RMS}$ \\
 2.25    &$-0.80$    &4.33      &2.25    &$-4.58$    &2.06\\
 2.35    &$-2.09$    &2.89      &2.35    &$-7.39$    &1.65\\
 2.45    &$-4.00$    &2.14      &2.45    &$-10.96$   &1.38\\
 2.55    &$-6.59$    &1.70      &2.55    &$-15.37$   &1.18\\
 2.65    &$-9.90$    &1.41      &2.65    &$-20.68$   &1.03\\
 \bottomrule[1pt]
\bottomrule[1pt]
\end{tabular}
\end{table}

Compared to the $DK$ and $D^*K$ systems, there includes a long range force from the one$-\pi-$exchange (OPE) contribution for the $D^*K^*$ systems. Our results also indicate that the OPE interaction plays an important role in forming the $D^*K^*$ molecular state with $I(J^P)=1(0^+)$, the reason is that we cannot find bound state solutions until the cutoff larger than 4.90 GeV after neglecting the OPE interaction.

In summary, the newly $T_{c\bar{s}}^{a0(++)}(2900)$ can be identified as the $D^*K^*$ molecule with $I(J^P)=1(0^+)$, both the OPE interaction and the $S-D$ wave mixing effects are very important in reproducing its mass position.

In our previous paper \cite{Chen:2016ypj}, we analyzed the $D^{(*)}K^{(*)}$ interactions with other different quantum number configurations by using the OBE model. Finally, we find that the isoscalar $D^{(*)}K^{(*)}$ systems with $J^P=0^+$, $1^+$, and $2^+$ are likely to be possible molecular candidates. However, for the isovector $D^{(*)}K^{(*)}$ systems with $1^+$ and $2^+$, their OBE effective potentials are much weaker, i.e., they are not strong enough to form hadronic molecular states.

\section{Predicting possible charmed-strange molecular pentaquarks from the $\Lambda_{c}K^{(*)}$ and $\Sigma_{c}K^{(*)}$ interactions}\label{sec3}

For the heavy baryon $Q(qq)$, the diquark $(qq)$ has the same color structure $\bar{3}_c$ with an anti-quark $(\bar{q})$ in the heavy meson $Q\bar q$. In this section, we extend to study the $\Lambda_{c}K^{(*)}$ and $\Sigma_{c}K^{(*)}$ interactions by using the OBE model, where we can search for possible charmed-strange molecular pentaquarks. The relevant Lagrangians describing the interactions between the heavy baryons and light mesons are constructed in terms of the heavy quark limit and chiral symmetry \cite{Liu:2011xc}, i.e.,
\begin{eqnarray}
\mathcal{L}_{\sigma} &=& l_B\langle \bar{\mathcal{B}}_{\bar{3}}\sigma\mathcal{B}_{\bar{3}}\rangle
      -l_S\langle\bar{\mathcal{B}}_6\sigma\mathcal{B}_6\rangle,\\
\mathcal{L}_{{P}} &=&
        i\frac{g_1}{2f_{\pi}}\varepsilon^{\mu\nu\lambda\kappa}v_{\kappa}\langle\bar{\mathcal{B}}_6
        \gamma_{\mu}\gamma_{\lambda}\partial_{\nu}{P}\mathcal{B}_6\rangle\nonumber\\
      &&-\sqrt{\frac{1}{3}}\frac{g_4}{f_{\pi}}\langle\bar{\mathcal{B}}_6\gamma^5
      \left(\gamma^{\mu}+v^{\mu}\right)\partial_{\mu}{P}\mathcal{B}_{\bar{3}}\rangle+h.c.,\\
\mathcal{L}_{{V}} &=& \frac{1}{\sqrt{2}}\beta_Bg_V\langle\bar{\mathcal{B}}_{\bar{3}}v\cdot{V}\mathcal{B}_{\bar{3}}\rangle
   -\frac{\beta_Sg_V}{\sqrt{2}}\langle\bar{\mathcal{B}}_6v\cdot{V}\mathcal{B}_6\rangle\nonumber\\
    &&-i\frac{\lambda g_V}{3\sqrt{2}}\langle\bar{\mathcal{B}}_6\gamma_{\mu}\gamma_{\nu}
    \left(\partial^{\mu} {V}^{\nu}-\partial^{\nu} {V}^{\mu}\right)
    \mathcal{B}_6\rangle\nonumber\\
    &&-\frac{\lambda_Ig_V}{\sqrt{6}}\varepsilon^{\mu\nu\lambda\kappa}v_{\mu}\langle \bar{\mathcal{B}}_6\gamma^5\gamma_{\nu}
        \left(\partial_{\lambda} {V}_{\kappa}-\partial_{\kappa} {V}_{\lambda}\right)\mathcal{B}_{\bar{3}}\rangle+h.c..\nonumber\\
\end{eqnarray}
The matrices $\mathcal{B}_{\bar{3}}$ and $\mathcal{B}_6$ denote as
\begin{eqnarray*}
\mathcal{B}_{\bar{3}} = \left(\begin{array}{ccc}
        0    &\Lambda_c^+     \\
        -\Lambda_c^+       &0
\end{array}\right),\quad\quad
\mathcal{B}_6 = \left(\begin{array}{ccc}
         \Sigma_c^{++}                  &\frac{\Sigma_c^{+}}{\sqrt{2}}\\
         \frac{\Sigma_c^{+}}{\sqrt{2}}      &\Sigma_c^{0}
\end{array}\right).
\end{eqnarray*}
The coupling constants in the above Lagrangians are taken from Ref. \cite{Liu:2011xc}, $l_S=-2l_B=7.3$, $g_1=(\sqrt{8}/3)g_1=1.0$, $\beta_Sg_V=-2\beta_Bg_V=12.0$, $\lambda_Sg_V=-2\sqrt{2}\lambda_Ig_V=19.2~ \text{GeV}^{-1}$.

\paragraph{\bf{The $\Lambda_cK^{(*)}$ systems.}} For this kind of system, the flavor wave function $|I,I_3\rangle$ can be expressed as $|1/2,1/2\rangle=|\Lambda_c^+K^{(*)+}\rangle$ and $|1/2,-1/2\rangle=|\Lambda_c^+K^{(*)0}\rangle$. When we consider the $S-D$ wave mixing effects, their spin-orbit wave functions are
\begin{eqnarray*}\left.\begin{array}{llll}
\Lambda_cK[J^P=1/2^-]: &|{}^2S_{1/2}\rangle,\\
\Lambda_cK^*[J^P=1/2^-]:   &|{}^{2}S_{1/2}\rangle,  &|{}^{4}D_{1/2}\rangle,\\
\Lambda_cK^*[J^P=3/2^-]:   &|{}^{4}S_{3/2}\rangle,  &|{}^{2}D_{3/2}\rangle,  &|{}^{4}D_{3/2}\rangle.
\end{array}\right.\end{eqnarray*}
With the same procedures, we can deduce the OBE effective potentials for the $\Lambda_cK^{(*)}$ systems as follows,
\begin{eqnarray}
V_{\Lambda_cK} &=& -l_Bg_{\sigma}Y(\Lambda,m_{\sigma},r)+\beta_Bg_Vg_{\omega KK}Y(\Lambda,m_{\omega},r),\\
V_{\Lambda_cK^*} &=& -l_Bg_{\sigma}(\bm{\epsilon}_2\cdot\bm{\epsilon}_4^{\dag})Y(\Lambda,m_{\sigma},r)\nonumber\\
    &&+\beta_Bg_Vg_{\omega K^*K^*}(\bm{\epsilon}_2\cdot\bm{\epsilon}_4^{\dag})Y(\Lambda,m_{\omega},r).
\end{eqnarray}
In above effective potentials, there only exists the intermediate range and short range forces from the one$-\sigma-$exchange and one$-\omega-$exchange interactions. For the single $S-$wave $\Lambda_cK$ system, we don't find bound solutions in the cutoff region $0.8\leq\Lambda\leq5.0$ GeV.

For the single $S-$wave $\Lambda_cK^*$ systems with $J^P=1/2^-$ and $3/2^-$, their OBE effective potentials are the same. In the cutoff region $0.8\leq\Lambda\leq5.0$ GeV, we cannot find bound state solutions, either. In the following, we further perform a coupled channel analysis on the $\Lambda_cK^*/\Sigma_cK^*$ interactions, in which the OBE effective potentials can be expressed as
\begin{eqnarray}
V_{\Lambda_cK^*}^{\text{C}} &=& \left(\begin{array}{cc}
V_{\Lambda_cK^*}    &V_{\Sigma_cK^*\to\Lambda_cK^*}\\
V_{\Lambda_cK^*\to\Sigma_cK^*}     &V_{\Sigma_cK^*}\end{array}\right),
\end{eqnarray}
with
\begin{eqnarray}
V_{\Lambda_cK^*\to\Sigma_cK^*} &=& -\frac{g_4g_{\pi K^*K^*}}{6f_{\pi}}
   \mathcal{Z}^{12}_{\Lambda_1,m_{\pi1}}
   -\frac{\lambda_Ig_Vg_{\rho K^*K^*}}{6\sqrt{2}m_{K^*}m_K}
    \mathcal{Z}^{\prime12}_{\Lambda_1,m_{\rho1}},\nonumber\\
V_{\Sigma_cK^*} &=& \frac{ l_Sg_{\sigma}}{2}\mathcal{Y}^{22}_{\Lambda ,m_{\sigma}}
  +\left[\frac{\beta_Sg_V}{8}+\frac{\lambda_Sg_V}{24m_{\Sigma_c}}\right]
   g_{\omega K^*K^*}\mathcal{Y}^{22}_{\Lambda,m_{\omega}}\nonumber\\
   &&+\left[\frac{\beta_Sg_V}{8}+\frac{\lambda_Sg_V}{24m_{\Sigma_c}}\right]
   \mathcal{G}(I)g_{\rho K^*K^*}\mathcal{Y}^{22}_{\Lambda,m_{\rho}}\nonumber\\
   &&-\frac{\lambda_Sg_V}{36m_{K^*}}\left[\mathcal{G}(I)g_{\rho K^*K^*}\mathcal{Z}^{\prime22}_{\Lambda,m_{\rho}}+g_{\omega K^*K^*}\mathcal{Z}^{\prime12}_{\Lambda,m_{\omega}}\right]\nonumber\\
   &&+\frac{g_1}{6\sqrt{2}f_{\pi}}\left[g_{\pi K^*K^*}\mathcal{G}(I)\mathcal{Z}^{22}_{\Lambda,m_{\pi}}
   +\frac{g_{\eta K^*K^*}}{3}\mathcal{Z}^{22}_{\Lambda,m_{\eta}}\right].\label{SKX}\nonumber\\
\end{eqnarray}
Here, $\Lambda_1^2 =\Lambda^2-q_1^2$, $m_{{\pi1}}^2=m_{\pi}^2-q_1^2$, $m_{\rho1}^2=m_{\rho}^2-q_1^2$ with $q_1 =
\frac{M_{\Sigma_c}^2-M_{\Lambda_c}^2}{2(M_{\Sigma_c}+M_{K^*})}$, and the definitions of several useful functions are
\begin{eqnarray}
\mathcal{Y}^{ij}_{\Lambda, m_a}&=&\mathcal{D}_{ij}Y(\Lambda,m_\sigma,r),\\
\mathcal{Z}^{ij}_{\Lambda, m_a}&=&\left(\mathcal{E}_{ij}\nabla^2+\mathcal{F}_{ij}r\frac{\partial}{\partial r}\frac{1}{r}\frac{\partial}{\partial r}\right)Y(\Lambda,m_a,r),\\
\mathcal{Z}^{\prime ij}_{\Lambda,
m_a}&=&\left(2\mathcal{E}_{ij}\nabla^2-\mathcal{F}_{ij}r\frac{\partial}{\partial
r}\frac{1}{r}\frac{\partial}{\partial r}\right)Y(\Lambda,m_a,r),
\end{eqnarray}
where $\mathcal{E}_{12}=i\bm{\sigma}\cdot(\bm{\epsilon}_2\times\bm{\epsilon}_4^{\dag})$, $\mathcal{F}_{12}=S(\hat{r},\bm{\sigma},i\bm{\epsilon}_2\times\bm{\epsilon}_4^{\dag})$, $\mathcal{D}_{22}=\bm{\epsilon}_2\cdot\bm{\epsilon}_4^{\dag}$, $\mathcal{E}_{22}=i\bm{\sigma}\cdot(\bm{\epsilon}_2\times\bm{\epsilon}_4^{\dag})$, and $\mathcal{F}_{22}=S(\hat{r},\bm{\sigma},i\bm{\epsilon}_2\times\bm{\epsilon}_4^{\dag})$. As a result, the numerical matrices for the above operators can be obtained by the spin-orbit wave functions for the discussed systems as
\begin{eqnarray}
i\bm{\sigma}\cdot(\bm{\epsilon}_2\times\bm{\epsilon}_4^{\dag})\mapsto
\left\{\begin{array}{ll}\left(\begin{array}{cc}-2  &1\\ 0 &1\end{array}\right),  &J^P=1/2^-\\
\left(\begin{array}{ccc}1  &0  &0\\0  &-2  &0\\ 0  &0  &1\end{array}\right),  &J^P=3/2^-\end{array}\right.\\
S(\hat{r},\bm{\sigma},i\bm{\epsilon}_2\times\bm{\epsilon}_4^{\dag})\mapsto
\left\{\begin{array}{ll}\left(\begin{array}{cc}0  &-\sqrt{2}\\ -\sqrt{2} &-2\end{array}\right),  &J^P=1/2^-\\
\left(\begin{array}{ccc}0  &1  &2\\ 1  &0  &-1\\ 2  &-1  &0\end{array}\right),  &J^P=3/2^-\end{array}\right.
\end{eqnarray}

\renewcommand\tabcolsep{0.2cm}
\renewcommand{\arraystretch}{1.7}
\begin{table*}[!hbtp]
\caption{The bound state solutions (the binding energy $E$, the root-mean-square radius $r_{RMS}$, and the probabilities $P_i(\%)$ for all the discussed channels) for the coupled $\Lambda_cK^*/\Sigma_cK^*$ systems with $I(J^P)=1/2(1/2^-)$ and $1/2(3/2^-)$. Here, $E$, $r_{RMS}$, and $\Lambda$ are in units of MeV, fm, and GeV, respectively. The dominant channels are labeled in a bold manner.}\label{num2}
\begin{tabular}{c|ccc|cccccc}
\toprule[1pt]
 $I(J^{P})$      &$\Lambda$    &$E$       &$r_{RMS}$
      &$\Lambda_cK^*({}^2S_{1/2})$   &$\Lambda_cK^*({}^{4}D_{1/2})$     &$\Sigma_cK^*({}^{2}S_{1/2})$      &$\Sigma_cK^*({}^{4}D_{1/2})$\\\midrule[1pt]
$1/2(1/2^-)$     &1.27   &$-2.71$   &1.76    &\bf{85.75}   &0.13    &14.11   &0.01\\
                 &1.28   &$-7.38$   &1.02    &\bf{78.09}   &0.19    &21.71   &0.02\\
                 &1.29   &$-12.86$  &0.77    &\bf{72.15}   &0.24    &27.59   &0.02\\
                 &1.30   &$-18.93$  &0.65    &\bf{67.46}   &0.27    &32.26   &0.02\\
                 &1.31   &$-25.47$  &0.58    &\bf{63.64}   &0.29    &36.05   &0.02 \\\midrule[1pt]
$I(J^{P})$      &$\Lambda$    &$E$       &$r_{RMS}$
      &$\Lambda_cK^*({}^4S_{3/2})$   &$\Lambda_cK^*({}^{2}D_{3/2})$     &$\Lambda_cK^*({}^{4}D_{3/2})$   &$\Sigma_cK^*({}^{4}S_{3/2})$      &$\Sigma_cK^*({}^{2}D_{3/2})$   &$\Sigma_cK^*({}^{4}D_{3/2})$
      \\\midrule[1pt]
$1/2(3/2^-)$  &2.71   &$-2.42$  &1.25  &29.02   &0.60   &6.18  &\bf{59.30}   &2.77 &2.12\\
              &2.72   &$-5.72$  &0.82  &26.72   &0.62   &6.35  &\bf{61.26}   &2.85 &2.20\\
              &2.73   &$-9.23$  &0.68  &25.29   &0.62   &6.45  &\bf{62.49}   &2.90 &2.24\\
              &2.74   &$-12.87$ &0.61  &24.29   &0.63   &6.51  &\bf{63.35}   &2.94 &2.28\\
              &2.75   &$-16.63$ &0.57  &23.54   &0.63   &6.55  &\bf{64.01}   &2.96 &2.30\\
\bottomrule[1pt]
\end{tabular}
\end{table*}

In Table \ref{num2}, we collect the bound state properties for the coupled $\Lambda_cK^*/\Sigma_cK^*$ systems with $I(J^P)=1/2(1/2^-)$ and $1/2(3/2^-)$. For the coupled $\Lambda_cK^*/\Sigma_cK^*$ system with $I(J^P)=1/2(1/2^-)$, we can obtain a loosely bound state with the binding energy $E=-2.71$ MeV and the RMS radius $r_{RMS}=1.76$ fm at cutoff $\Lambda=1.27$ GeV, and the dominant channel is the $\Lambda_cK^*({}^2S_{1/2})$. After increasing the cutoff $\Lambda$, this bound state binds deeper and deeper, the contribution from the $\Sigma_cK^*({{}^2S_{1/2}})$ channel becomes more and more important. Compared to the results in the single channel analysis, we can conclude that the coupled channel effects play an important role in forming this bound state. If we adopt the cutoff value of the deuteron~\cite{Tornqvist:1993ng,Tornqvist:1993vu}, the coupled $\Lambda_cK^*/\Sigma_cK^*$ systems with $I(J^P)=1/2(1/2^-)$ can be the prime hadronic molecular candidate.

For the coupled $\Lambda_cK^*/\Sigma_cK^*$ system with $1/2(3/2^-)$, the loosely bound state solutions appear at the cutoff $\Lambda$ around 2.71 GeV, which is much larger than the cutoff value in the $1/2(1/2^-)$ bound state. Therefore, we find that the OBE effective potentials for the coupled $\Lambda_cK^*/\Sigma_cK^*$ system with $1/2(3/2^-)$ is much weaker attractive than those in the coupled $\Lambda_cK^*/\Sigma_cK^*$ system with $1/2(1/2^-)$. As shown in Table \ref{num2}, the $\Sigma_cK^*({{}^4S_{3/2}})$ is the dominant channel for the coupled $\Lambda_cK^*/\Sigma_cK^*$ system with $1/2(3/2^-)$. In general, the size $R$ of the $S-$wave loosely bound state can be estimated by the reduced mass $\mu$, the binding energy $E$, and the mass thresholds for the lowest channel $M_{\text{L}}$ and the dominant channel $M_{\text{D}}$ in the coupled channel analysis, i.e., $R\sim1/\sqrt{2\mu |E+M_{\text{L}}-M_{\text{D}}|}$ \cite{Close:2003sg,Chen:2017xat}. According to this approximation relation, the RMS radii for the coupled $\Lambda_cK^*/\Sigma_cK^*$ bound state with $1/2(3/2^-)$ are a little smaller than the size of the coupled $\Lambda_cK^*/\Sigma_cK^*$ system with $1/2(1/2^-)$ with the same binding energy. If we still adopt the experience of the deuteron~\cite{Tornqvist:1993ng,Tornqvist:1993vu}, the coupled $\Lambda_cK^*/\Sigma_cK^*$ system with $1/2(3/2^-)$ cannot be a good hadronic molecular candidate.

\paragraph{\bf{The $\Sigma_cK^{(*)}$ systems.}} The spin-orbit wave functions are same with the $\Lambda_cK^{(*)}$ systems. For the flavor wave functions, they can be written as
\begin{eqnarray*}&&\begin{array}{c}
\left|\frac{1}{2},\frac{1}{2}\right\rangle =
     \sqrt{\frac{2}{3}}\left|\Sigma_c^{++}{K}^{(*)0}\right\rangle
     -\frac{1}{\sqrt{3}}\left|\Sigma_c^{+}{K}^{(*)+}\right\rangle,\\
\left|\frac{1}{2},-\frac{1}{2}\right\rangle =
     \frac{1}{\sqrt{3}}\left|\Sigma_c^{+}{K}^{(*)0}\right\rangle
     -\sqrt{\frac{2}{3}}\left|\Sigma_c^{0}{K}^{(*)+}\right\rangle,
     \end{array}\\
&&\begin{array}{l}
\left|\frac{3}{2},\frac{3}{2}\right\rangle = \left|\Sigma_c^{++}{K}^{(*)+}\right\rangle,\\
\left|\frac{3}{2},\frac{1}{2}\right\rangle =
     \frac{1}{\sqrt{3}}\left|\Sigma_c^{++}{K}^{(*)0}\right\rangle+
     \sqrt{\frac{2}{3}}\left|\Sigma_c^{+}{K}^{(*)+}\right\rangle,\\
\left|\frac{3}{2},-\frac{1}{2}\right\rangle =\sqrt{\frac{2}{3}}
    \left|\Sigma_c^{+}{K}^{(*)0}\right\rangle
    +\frac{1}{\sqrt{3}}\left|\Sigma_c^{0}{K}^{(*)+}\right\rangle,\\
\left|\frac{3}{2},-\frac{3}{2}\right\rangle =
     \left|\Sigma_c^{0}{K}^{(*)0}\right\rangle.
     \end{array}
\end{eqnarray*}
The OBE effective potentials for the single $\Sigma_cK$ can be deduced as
\begin{eqnarray}
V_{\Sigma_cK} &=& \frac{1}{2}l_Sg_{\sigma}Y(\Lambda,m_{\sigma},r)
    -\frac{\mathcal{G}(I)}{2}\beta_Sg_Vg_{\rho KK}Y(\Lambda,m_{\rho},r)\nonumber\\
    &&+\frac{\mathcal{G}(I)}{6m_{\Sigma_c}}\lambda_Sg_Vg_{\rho KK}\nabla^2Y(\Lambda,m_{\rho},r)\nonumber\\
    &&-\frac{1}{2}\beta_Sg_Vg_{\omega KK}Y(\Lambda,m_{\omega},r)\nonumber\\
    &&+\frac{1}{6m_{\Sigma_c}}\lambda_Sg_Vg_{\omega KK}\nabla^2Y(\Lambda,m_{\omega},r).
\end{eqnarray}
Here, there exists the additional $\rho-$exchange interaction in comparison with the $\Lambda_cK$ system, and it provides the attractive and repulsive forces for the $\Sigma_cK$ system with $I=1/2$ and $3/2$, respectively. Our results show that there exist no bound state solutions for the iso-quartet $\Sigma_cK$ system. For the iso-doublet $\Sigma_cK$ system, as presented in Table \ref{num3}, we can obtain the loosely bound state solutions as the cutoff $\Lambda$ is larger than 3.40 GeV.

Here, we notice that the mass gap between the $\Sigma_cK$ and $\Lambda_cK^*$ systems is small, and the $\pi-$exchange interactions are also allowed for the $\Lambda_cK^*\to\Sigma_cK$ process, therefore, the coupled channel effects may be important in the coupled $\Sigma_cK/\Lambda_cK^*$ channel analysis. The corresponding OBE effective potentials can be expressed as
\begin{eqnarray}
V_{\Sigma_cK}^{\text{C}} &=& \left(\begin{array}{cc}
V_{\Sigma_cK}    &V_{\Lambda_cK^*\to\Sigma_cK}\\
V_{\Sigma_cK\to\Lambda_cK^*}     &V_{\Lambda_cK^*}\end{array}\right),
\end{eqnarray}
with
\begin{eqnarray}
V_{\Lambda_cK^*\to\Sigma_cK} &=& -\frac{g_4g_{\pi KK^*}}{3f_{\pi}}
    \left[\bm{\sigma}\cdot\bm{\epsilon_2}\nabla^2\right.\nonumber\\
    &&\left.+S(\hat{r},\bm{\sigma},\bm{\epsilon_2})r\frac{\partial}{\partial r}\frac{1}{r}\frac{\partial}{\partial r}\right]U(\Lambda_0,m_{\pi0},r)\nonumber\\
    &&-\frac{\lambda_Ig_Vg_{\rho KK^*}}{12\sqrt{2}}\sqrt{\frac{m_{\Lambda_c}m_{K^*}}{m_{\Sigma_c}m_K}}
    \left[2\bm{\sigma}\cdot\bm{\epsilon_2}\nabla^2\right.\nonumber\\
    &&\left.-S(\hat{r},\bm{\sigma},\bm{\epsilon_2})r\frac{\partial}{\partial r}\frac{1}{r}\frac{\partial}{\partial r}\right]Y(\Lambda_0,m_{\rho0},r).
\end{eqnarray}
Here, $q_0=\frac{M_{\Lambda_c}^2+M_{K}^2-M_{\Sigma_c}^2-M_{K^*}^2}{2(M_{\Sigma_c}+M_{K})}$, $\Lambda_0^2=\Lambda^2-q_0^2$, $m_{\pi0}^2=q_0^2-m_{\pi}^2$, $m_{\rho0}^2=m_{\rho}^2-q_0^2$. We define
\begin{eqnarray*}
U(\Lambda,m,{r}) &=& \frac{1}{4\pi r}\left(\cos(mr)-e^{-\Lambda r}\right)-\frac{\Lambda^2+m^2}{8\pi \Lambda}e^{-\Lambda r}.
\end{eqnarray*}
After considering the $S-D$ wave mixing effects, the matrix elements for the spin-spin interaction and tensor force operators read as $\langle\bm{\sigma}\cdot\bm{\epsilon_2}\rangle\mapsto (\begin{array}{cc}\sqrt{3} &0\end{array} )$ and $\langle S(\hat{r},\bm{\sigma},\bm{\epsilon_2})\rangle\mapsto (\begin{array}{cc}0 &-\sqrt{6}\end{array} )$, respectively.

\renewcommand\tabcolsep{0.15cm}
\renewcommand{\arraystretch}{1.6}
\begin{table}[!hbtp]
\caption{The bound state solutions (the binding energy $E$, the root-mean-square radius $r_{RMS}$, and the probabilities $P_i(\%)$ for all the discussed channels) for the single $\Sigma_cK$ and coupled $\Sigma_cK/\Lambda_cK^*$ systems with $I(J^P)=1/2(1/2^-)$. Here, $E$, $r_{RMS}$, and $\Lambda$ are in units of MeV, fm, and GeV, respectively. Notations $S_1$, $S_2$, and $D_2$ stand for the $\Sigma_cK({}^2S_{1/2})$, $\Lambda_cK^*({}^2S_{1/2})$, $\Lambda_cK^*({}^4D_{1/2})$ channels, respectively.} \label{num3}
\begin{tabular}{ccc|cccc}
\toprule[1pt]\toprule[1pt]
\multicolumn{3}{c|}{Single channel}    &\multicolumn{4}{c}{Coupled channel}\\\midrule[1pt]
$\Lambda$   &$E$ &$r_{RMS}$
 &$\Lambda$    &$E$  &$r_{RMS}$   &$P_i(S_1,~S_2,~D_2)$\\
 3.40    &$-0.18$    &6.39      &1.37   &$-0.29$    &6.18    &(99.00,~0.62,~0.38)\\
 3.80    &$-1.79$    &3.73      &1.42   &$-2.08$    &3.57    &(98.37~,1.04,~0.59)\\
 4.20    &$-4.50$    &2.47      &1.47   &$-5.55$    &2.29    &(97.63,~1.56,~0.80)\\
 4.60    &$-8.17$    &1.87      &1.52   &$-11.00$   &1.67    &(96.85,~2.14,~1.00)\\
 5.00    &$-12.72$   &1.52      &1.57   &$-18.67$   &1.31    &(96.05,~2.76,~1.19)\\
 \bottomrule[1pt]
\bottomrule[1pt]
\end{tabular}
\end{table}

In Table \ref{num3}, we collect the bound state solutions (the binding energy $E$, the root-mean-square radius $r_{RMS}$, and the probabilities $P_i(\%)$ for all the discussed channels) for the coupled $\Sigma_cK/\Lambda_cK^*$ system with $I(J^P)=1/2(1/2^-)$. When the cutoff is taken around 1.40 GeV, the binding energy is around several MeV, the RMS radius is several fm. The $\Sigma_cK({}^2S_{1/2})$ channel is the dominant channel with the probability around 98\%. Compared to the bound state properties in the single channel case, the cutoff here is very close to the value $\Lambda\sim1.00$ GeV as in the deuteron. Therefore, we conclude that the coupled $\Sigma_cK/\Lambda_cK^*$ systems with $I(J^P)=1/2(1/2^-)$ can be prime molecular candidate, and the coupled channel effects play an important role.

\renewcommand\tabcolsep{0.1cm}
\renewcommand{\arraystretch}{1.6}
\begin{table}[!hbtp]
\caption{The $\Lambda$ dependence of the obtained bound-state solutions (the binding energy $E$ and the root-mean-square radius $r_{RMS}$) for the $\Sigma_c K^*$ systems. Here, $E$, $r_{RMS}$, and $\Lambda$ are in units of MeV, fm, and GeV, respectively.}\label{num4}
\begin{tabular}{cccc|cccc}
\toprule[1pt]\toprule[1pt]
$I(J^P)$ &$\Lambda$   &$E$ &$r_{RMS}$
 &$I(J^P)$  &$\Lambda$    &$E$  &$r_{RMS}$ \\\hline
 $1/2(1/2^-)$   &1.12     &$-0.36$   &5.33
 &$1/2(3/2^-)$   &2.90     &$-0.21$   &6.12\\
                &1.16     &$-1.55$   &3.22
                &&3.10     &$-1.48$   &3.47\\
                &1.20     &$-3.52$   &2.22
                &&3.30     &$-4.08$   &2.22\\
                &1.24     &$-6.32$   &1.70
                &&3.50     &$-8.36$   &1.62\\
                &1.28     &$-10.01$  &1.39
                &&3.70     &$-14.70$  &1.27\\\hline
 $3/2(1/2^-)$   &3.75     &$-0.25$   &5.96
 &$3/2(3/2^-)$  &1.45     &$-0.46$   &5.12\\
                &4.00     &$-0.80$   &4.44
                &&1.53     &$-2.66$   &2.57\\
                &4.25     &$-1.60$   &3.38
                &&1.61     &$-6.66$   &1.70\\
                &4.50     &$-2.63$   &2.71
                &&1.69     &$-12.52$  &1.28\\
                &4.75     &$-3.90$   &2.28
                &&1.77     &$-20.33$  &1.04\\
 \bottomrule[1pt]
\bottomrule[1pt]
\end{tabular}
\end{table}

For the $\Sigma_cK^*$ systems, the isospin and spin-parity configurations $I(J^P)$ include $1/2(1/2^-)$, $1/2(3/2^-)$, $3/2(1/2^-)$, and $3/2(3/2^-)$. By adopting the OBE effective potentials in Eq. (\ref{SKX}), we can obtain loosely bound state solutions for all the discussed quantum number configurations in the cutoff range $0.80\leq\Lambda\leq5.00$ GeV. As shown in Table \ref{num4}, we find
\begin{enumerate}
  \item For the $\Sigma_cK^*$ systems with $1/2(1/2^-)$ and $3/2(3/2^-)$, the reasonable loosely bound state solutions (the binding energy is around several to several tens MeV and the RMS radius is around several fm.) appear at cutoff around 1.0 GeV, which is comparable to the value in the deuteron. These two states can be good hadronic molecular candidates.
  \item For the $\Sigma_cK^*$ systems with $1/2(3/2^-)$ and $3/2(1/2^-)$, we can find the loosely bound state solutions as the cutoff is around or even much larger than 3.00 GeV. These cutoff values are too far away from the empirical value for the deuteron. We conclude that they two are less likely to be good molecular candidates.
\end{enumerate}

In conclusion, our results can predict four possible charmed-strange molecular pentaquarks, the coupled $\Lambda_cK^*/\Sigma_cK^*$ molecular state with $I(J^P)=1/2(1/2^-)$, the coupled $\Sigma_cK/\Lambda_cK^*$ molecular state with $I(J^P)=1/2(1/2^-)$, and the single $\Sigma_cK^*$ state with $I(J^P)=1/2(3/2^-)$ and $3/2(3/2^-)$. In these predictions, the coupled channel effects are very important. Based on the conservation of the quantum numbers and the limit of the phase space, we can summary the strong decay channels for these predicted charmed-strange pentaquark molecules, i.e.,
\begin{eqnarray}
\Sigma_cK/\Lambda_cK^*[1/2(1/2^-)]\to && \left\{D_sN, \Lambda_cK\right\},\nonumber\\
\Lambda_cK^*/\Sigma_cK^*[1/2(1/2^-)]\to && \left\{D_s^{(*)}N, \Lambda_cK, \Sigma_cK\right\},\nonumber\\
\Sigma_cK^*[1/2(1/2^-)]\to && \left\{D_s^{(*)}N, \Lambda_cK^{(*)}, \Sigma_cK\right\},\nonumber\\
\Sigma_cK^*[3/2(3/2^-)]\to && \left\{D_s^{(*)}\Delta, \Sigma_c^*K\right\}.\nonumber
\end{eqnarray}

In Ref. \cite{An:2022vtg}, the authors systematically studied the mass spectrum and the decay properties of the charmed-strange pentaquark system $c\bar s qqq$ and suggested experiments to search them in the $b-$hadron decays.

\section{Summary}\label{sec4}

The study of the exotic states is an important and interesting issue in the hadron physics. Very recently, the LHCb collaboration reported the observation of the $T_{c\bar s}^{a0(++)}$ in the $D_s^+\pi^{-(+)}$ invariant mass spectrums. The observations of the newly $T_{c\bar s}^{a0(++)}$ again support the existence of the multiquark states as their decay modes. Since the masses of the newly $T_{c\bar s}^{a0(++)}$ are very close to the $D^*K^*$ threshold, in this work, we check the possibility of the newly $T_{c\bar s}^{a0(++)}$ as the isovector $D^*K^*$ molecular state with $J^P=0^+$. We study the $D^{(*)}K^{(*)}$ interactions by adopting the OBE effective potentials and considering the $S-D$ wave mixing effects. In order to estimate the cutoff value for the $D^{(*)}K^{(*)}$ interactions, we first reproduce the masses of the $D_{s0}(2317)$ and $D_{s1}(2460)$ in the $DK[0(0^+)]$ and $D^*K[0(1^+)]$ molecular picture, respectively. With the same parameters input, our results show that the newly observed $T_{c\bar s}^{a0(++)}$ can be explained as the $D^*K^*$ molecular state with $I(J^P)=1(0^+)$. The OPE interactions play a very important role in the formation of this bound state, and the $S-D$ wave mixing effects are also helpful.

After that, we extend our investigation on the $\Lambda_cK^{(*)}$ and $\Sigma_cK^{(*)}$ interactions with the same model. Our results indicate the single $\Sigma_cK^*$ state with $I(J^P)=1/2(1/2^-)$ and $3/2(3/2^-)$ can be good charmed-strange molecular candidates. When we further consider the coupled channel effects, we can predict another two possible charmed-strange molecular states, i.e., the coupled $\Lambda_cK^*/\Sigma_cK^*$ molecular state with $I(J^P)=1/2(1/2^-)$, the coupled $\Sigma_cK/\Lambda_cK^*$ molecular state with $I(J^P)=1/2(1/2^-)$, where the dominant channels are the $\Lambda_cK^*({}^2S_{1/2})$ and $\Sigma_cK({}^2S_{1/2})$, respectively, which means coupled channel effects play the key role in binding these two possible molecular candidates.

\section*{ACKNOWLEDGMENTS}

R. C. is supported by the Xiaoxiang Scholars Programme of Hunan Normal University.

\bibliography{ref}

\end{document}